%% file: main.tex
\documentclass[manuscript,screen]{acmart}
\AtBeginDocument{%
  }

\input{config}

\begin{document}
\title{QSPE: Enumerating Skeletal Quantum Programs for Quantum Library Testing}

\input{sec/0-abs}

\author{Jiaming Ye}
\authornote{Jiaming Ye is with the Manufacturing Industry Chain Collaboration Industrial Software Key Laboratory of Sichuan Province.}
\email{yejiaming@swjtu.edu.cn}
\affiliation{%
  \institution{Southwest Jiaotong University}
  \country{China}
}
\author{Fuyuan Zhang}
\authornote{Fuyuan Zhang is with the State Key Laboratory of Blockchain and Data Security.}
\affiliation{%
  \institution{Zhejiang University}
  \country{China}
}
\author{Shangzhou Xia}
\affiliation{%
  \institution{Kyushu University}
  \country{Japan}
}
\author{Xiaoyu Guo}
\affiliation{%
  \institution{Kyushu University}
  \country{Japan}
}
\author{Xiongfei Wu}
\affiliation{%
  \institution{University of Luxembourg}
  \country{Luxembourg}
}
\author{Jianjun Zhao}
\affiliation{%
  \institution{Kyushu University}
  \country{Japan}
}
\author{Yinxing Xue}
\affiliation{%
  \institution{Institute of AI for Industries}
  \country{China}
}

\maketitle

\input{sec/1-intro}
\input{sec/2-background}

\input{sec/3-workflow}
\input{sec/4-qspe}
\input{sec/5-exp}
\input{sec/6-ending}

\newpage
\bibliographystyle{ACM-Reference-Format}
\bibliography{mybib}

\end{document}

%% file: config.tex
\usepackage{tikz}
\usepackage{amsmath}
\usepackage{url}
\usepackage{balance}
\usepackage{listings}
\usepackage{calc}
\usepackage{bbding}
\usepackage{pifont}
\usepackage{wasysym}
\usepackage{xcolor}
\definecolor{mykeyword}{rgb}{0,0,0.6}  
\definecolor{mystring}{rgb}{0.6,0,0}   
\usepackage{graphicx}
\usepackage{caption,subcaption}
\usepackage{enumitem}
\usepackage{supertabular}
\usepackage{colortbl}
\usepackage{microtype}
\usepackage{fancyvrb}
\usepackage{multicol}
\usepackage{rotating}
\usepackage{enumitem}
\usepackage{algorithmic}
\usepackage{xspace}
\usepackage{braket}
\usepackage{booktabs}
\usepackage[most]{tcolorbox}
\usepackage{minted}
\tcbuselibrary{minted}

\usepackage[ruled,vlined,linesnumbered]{algorithm2e}
\usepackage{tikz}
\usepackage{mdframed}
\usepackage{multirow}
\usepackage{bigstrut}
\usepackage{mathtools}
\usepackage{mathpartir}
\usepackage{comment}
\usepackage{pifont}

\definecolor{myblue}{RGB}{0,32,96}

\newcolumntype{P}[1]{>{\raggedright\arraybackslash}p{#1}}

\newcommand{\genstirlingII}[3]{%
  \genfrac{\{}{\}}{0pt}{#1}{#2}{#3}%
}
\newcommand{\stirlingII}[2]{\genstirlingII{}{#1}{#2}}

\definecolor{todocolor}{rgb}{0.9,0.1,0.1}
\definecolor{ylcolor}{rgb}{0.7,0.7,0.3}

\usepackage{amssymb}    

%% file: sec/0-abs.tex
\begin{abstract}
The rapid advancement of quantum computing has led to the development of various quantum libraries, empowering compilation, simulation, and hardware backend interfaces. However, ensuring the correctness of these libraries remains a fundamental challenge due to the lack of mature testing methodologies. The state-of-the-art tools often rely on domain-specific configurations and expert knowledge, which limits their accessibility and scalability in practice. Furthermore, although these tools demonstrate strong performance, they adopt measurement-based for output validation in testing, which makes them produce false positive reports.

To alleviate these limitations, we propose QSPE, a practical approach that follows the differential testing principle and extends the existing approach, SPE, for quantum libraries. QSPE is fully automated, requiring no pre-set configurations or domain expertise, and can effectively generate a large set of diverse program variants that comprehensively explore the quantum compilation space. To mitigate the possible false positive reports, we propose statevector-based validation as an alternative to measurement-based validation.
In our experiments, the QSPE approach demonstrates remarkable effectiveness in generating 22,770 program variants across multiple quantum computing platforms. By avoiding $\alpha$-equivalence at the quantum and classical program wise, QSPE can reduce redundant generation and save more than 90\% of execution cost. Finally, the statevector-based validation method assists QSPE to reduce false alarms and effectively detects 708 miscompilations across multiple quantum libraries. Notably, 81 of the discovered bugs have been officially approved and acknowledged by the Qiskit development team, demonstrating the practical impact of our approach. 

\end{abstract}

%% file: sec/1-intro.tex
\section{Introduction}
\label{sec:1 intro}

Quantum software engineering has rapidly emerged as a pivotal research domain in recent years~\cite{murillo2025quantum}. This advancement is based on the development of quantum programming libraries such as Qiskit~\cite{wille2019ibm} and Cirq~\cite{isakov2021simulations}, which offer intuitive, Python-based interfaces for constructing and simulating quantum circuits. In particular, libraries provide a new programming paradigm: quantum-classical hybrid programs, where quantum modules are embedded in the classical language~\cite{serrano2022quantum, perez2021software}. This new paradigm has significantly reduced the entry barrier for quantum programming and accelerated the growth of the field.


However, the integration of quantum and classical components is prone to introduce bugs within quantum software stacks. Such bugs may arise from either the classical language logic or the quantum components themselves, potentially leading to miscompilations in downstream applications or even crashes during the compilation process. However, according to the state-of-the-art study~\cite{zhao2021bugs4q}, bugs in quantum libraries usually remain unfixed for years, highlighting the long-term needs of maintaining the reliability of quantum libraries. To address this, researchers have developed various testing approaches for testing quantum libraries. For example, QDiff~\cite{wang2021qdiff} uses gate transformation on equivalent programs to detect compiler bugs. MorphQ~\cite{paltenghi2023morphq} applies metamorphic testing to verify library behavior against predefined relations. More recently, UPBEAT~\cite{hu2024upbeat} has generated test cases by analyzing library API documentation and implementation.

Despite their contributions, the aforementioned approaches are proposed for particular testing scenarios. QDiff and MorphQ rely on pre-defined equivalence gate transformation rules or metamorphic rules, and UPBEAT requires specific API documentation and access to the source code. Furthermore, although they exhibit impressive effectiveness in testing, their performance may be limited as they adopt measurement sampling results for output validation (i.e., measurement-based validation), which can lead to false-positive reports. To support broader testing scenarios in quantum programs and improve the limitations of measurement-based validation, we propose QSPE, a practical approach that requires fewer input settings and can be applied to arbitrary quantum programs without pre-defined rules. Furthermore, we propose to enhance QSPE with statevector-based validation, which offers a robust alternative to measurement-based validation methods.

\begin{figure}[!tb]
\centering
\includegraphics[width=0.8\textwidth]{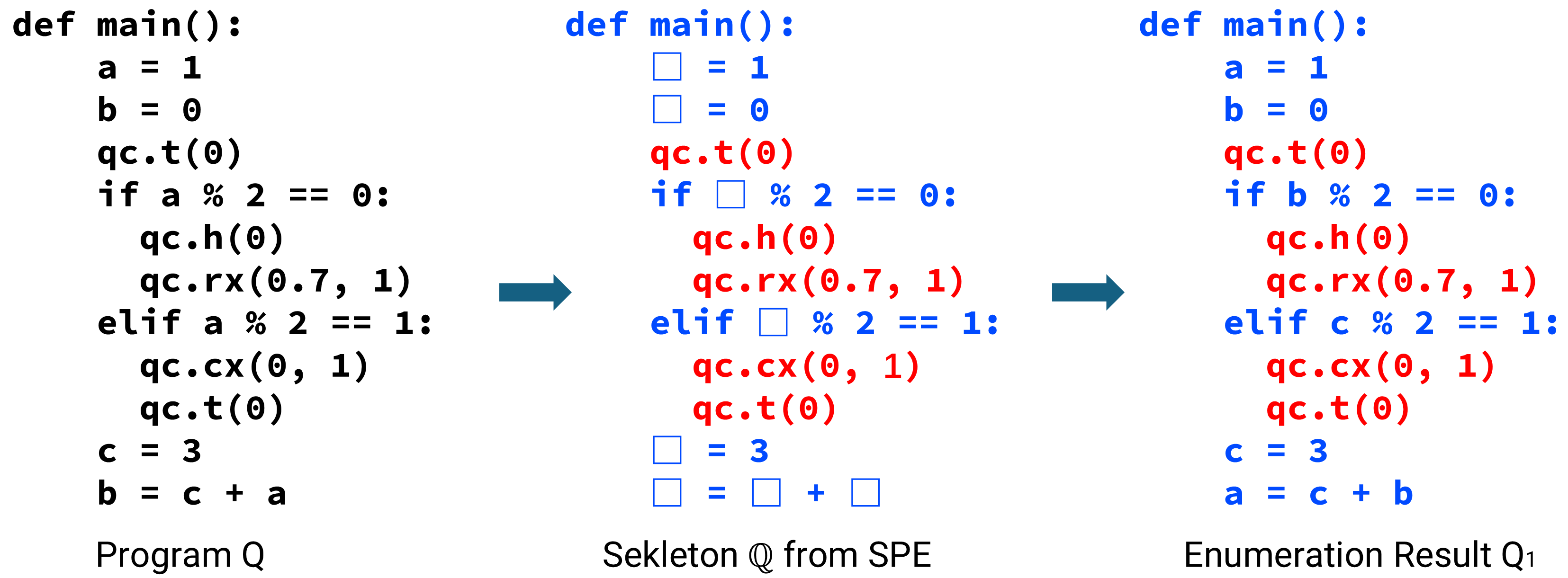}
\caption{A quantum program example $Q$, its skeleton $\mathbb{Q}$ by applying SPE, and one enumeration result $Q_1$. The quantum code, highlighted in red color, is overlooked by the SPE.}
\label{fig:spe}
\end{figure}

\textbf{Q}uantum \textbf{S}keletal \textbf{P}rogram \textbf{E}numeration is designed upon Skeletal Program Enumeration (SPE)~\cite{zhang2017skeletal}, an effective technique for generating a large number of test variants for testing C compilers. Although SPE is effective for C programs, it cannot handle quantum program code in practice, as shown in~\autoref{fig:spe}. The quantum components (highlighted in red) cannot be handled by the SPE method. In fact, as reported by~\cite{hidary2021quantum}, nearly 45\% of the code in quantum programs consists of quantum components. Therefore, it is challenging to apply the SPE approach directly to quantum library testing. Seeing this, QSPE bridges this gap by extending SPE to support the unique characteristics and requirements of quantum library testing.

Compared with existing approaches, QSPE is enhanced by: 1) adopting statevector-based validation method, which eliminates the incorrect assessments common in measurement-based approaches like QDiff; 2) re-designing the skeletal program model for the quantum domain by incorporating rotation angles and target qubits as key components of the variants, to enable rapid generation of numerous semantically rich quantum programs without requiring domain expertise or prior knowledge of quantum program structure; and 3) equipping QSPE with quantum-$\alpha$-equivalence filtering to reduce redundant test cases that exploit the same control and data dependencies, enabling QSPE to scale to large test requirements and different quantum libraries.
 
Our experimental results demonstrate the effectiveness of this approach. From 20 seed programs, QSPE generated 22,700 unique variants and successfully identified 708 miscompilations in two quantum libraries, including Qiskit and Pytket. In particular, 81 of the bugs discovered in Qiskit have been officially acknowledged and approved by its development team. We are actively working on the remaining 627 bugs, communicating with the Pytket developers to identify the root causes.

\textbf{\textit{Contributions.}} Our main contributions can be summarized as:

\begin{enumerate}
    \item We extend the prior SPE approach to support quantum library testing by addressing challenges between platforms and programming paradigms.
    \item We present the first report on the limitations of the measurement-based validation method, which is widely adopted in the state-of-the-art tools. As an alternative, we propose using the statevector-based validation method to verify its effectiveness.
    \item We propose the quantum-$\alpha$-equivalent. By filtering out equivalent programs, QSPE can reduce more than 90\% of the redundant programs, which largely saves test costs.
    \item To facilitate future research and community development, we open our enumeration results and make them a benchmark. The benchmark is publicly available at
    \begin{center}
    \url{https://sites.google.com/view/qspe/}
    \end{center}
    \item In experiments, QSPE effectively detects 708 miscompilations in the Qiskit and Pytket libraries. 81 of the 708 miscompilations have been approved by the developers in Qiskit. 
\end{enumerate}

\textbf{\textit{Paper organization.}} The rest of the paper is organized as follows. Section~\ref{sec:2 background} introduces quantum backgrounds that are closely related to our work. Section~\ref{sec:3 approach} illustrates the overall workflow of our approach, and Section~\ref{sec:4 qspe} defines the QSPE algorithm and the quantum-non-$\alpha$-equivalent. In Section~\ref{sec:evaluation}, we present our experimental results and discuss the findings. Finally, Section~\ref{sec:work} surveys related work and Section~\ref{sec:conclusion} concludes.

%% file: sec/2-background.tex
\section{Background}
\label{sec:2 background}

In this section, we introduce the background and concepts related to this work.

\begin{figure*}[!tb]
\centering
\includegraphics[width=\textwidth]{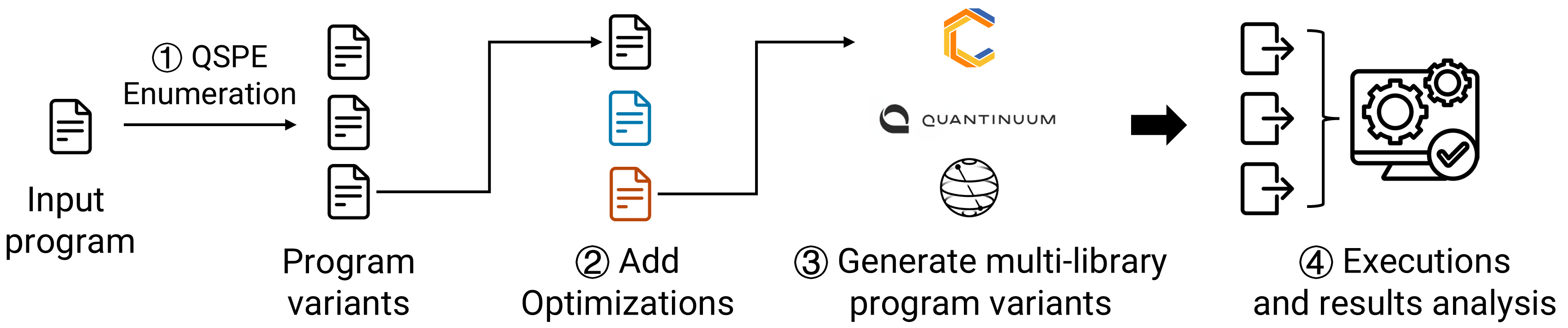}
\caption{The overall workflow of our approach.}
\label{fig:workflow}
\end{figure*}

\subsection{Qubit, Gates, Circuit, and Quantum Program}

A quantum bit (or qubit) is the most basic unit in quantum computation. The qubit is the quantum version of the classical bit in classical computers, but with quantum features. 
In quantum programs, the operations are represented by quantum logic gates. The gates provide various functions, including rotating a qubit to an arbitrary angle, assigning superposition to qubits, and creating controlled connections between qubits, among others. Some gates apply to only one qubit, while others apply to multiple qubits. The quantum circuit is the most basic unit for grouping and managing qubits and quantum gates. A quantum circuit is a model for quantum computation~\cite{quantum-book}, in which the qubits, gates, and measurements are organized in a sequence. The quantum program combines the above quantum components and can further include the classical components. The development of modern quantum libraries enables users to manipulate quantum circuits using Python, thus facilitating the integration of the quantum and classical components in quantum programs. 


\subsection{Statevector}

The statevector in a quantum program represents the complete quantum state of a quantum system as a complex-valued vector in the Hilbert space. For an n-qubit quantum system the statevector $\ket{\psi}$ is defined as:
\begin{equation*}
    \ket{\psi} = \sum_{\ket{x}\in B}{} A_{x} \ket{x}
\end{equation*}
where $A_{x}$ is the complex probability amplitudes, and the normalization condition $\sum_{\ket{x}\in B}{} \lvert A_{x} \rvert^{2} = 1$ must be satisfied.

In quantum programs, the statevector is computationally represented as an array of $2^n$ complex numbers, where each array index corresponds to a specific basis state $x$. The amplitude at each index determines the probability of measuring the system in that particular computational basis state, with the measurement probability given by $ \lvert A_{x} \rvert^{2}$. The statevector is updated through unitary transformations, where each quantum gate performs a linear, norm-preserving operation on the quantum state.

From a quantum program developer's perspective, the statevector serves as the fundamental data structure for quantum state representation in simulators and quantum virtual machines. The statevector enables precise tracking of quantum superposition and entanglement throughout program execution.

\subsection{Measuring the Output of a Quantum Program}

To read out the value of a qubit, a \emph{measurement} operation can be added to the qubit at the end of the program. The measurement operation causes the quantum state to collapse. Due to the fundamental properties of quantum mechanics, a qubit state $\ket{\psi}$ exists in a superposition of two basis states $\ket{0}$ and $\ket{1}$. After measurement, the state collapses to a definite value of either 0 or 1, with probabilities determined by the amplitude of the superposition. During this collapse, the information about the quantum state, such as phase and magnitude, is irreversibly lost, leaving only the classical measurement result. 

In the general case of an $n$-qubit program, we typically add $n$ measurement operations, one for each qubit, at the end of the program. After measurement, the results are read out as an $n$-bit binary number. Due to the probabilistic nature of the output values, a single measurement is usually insufficient to draw meaningful conclusions. To address this issue, a solution is to repeat the measurement process, which is commonly referred to as repeated $m$ \emph{measurements} (or \emph{samplings}). According to the law of large numbers, with an increasing number of samplings, the frequency of observed values is likely to converge to the actual probability distribution. For a quantum program, if the number of samplings is sufficiently large, the distribution of output values should be close to the actual distribution. However, determining the minimum number of samplings required by a quantum program is challenging because of the infeasibility of predicting the result distributions of quantum programs.

%% file: sec/3-workflow.tex
\section{Approach}
\label{sec:3 approach}
Our approach, QSPE, targets three major quantum libraries: Cirq, Qiskit, and Pytket. We selected these three libraries because Cirq and Qiskit are among the most widely used quantum programming frameworks, while Pytket has emerged as a rapidly growing library in recent years. The overall workflow, illustrated in~\autoref{fig:workflow}, consists of four main stages which we will detail in this section.

\subsection{Step 1: QSPE Enumeration}

The first stage of our workflow involves generating a large corpus of quantum program variants from a small set of initial seed programs. We begin with a seed pool of hybrid quantum-classical programs. Note that previous benchmarks, including MQT Bench~\cite{quetschlich2023mqt} and VeriQBench~\cite{chen2022veriqbench}, focus on datasets that comprise implementations of standalone quantum algorithms. In contrast, our study targets the hybrid quantum–classical paradigm, for which currently there is no suitable dataset. Therefore, we generate a new dataset with the help of large language models (LLMs). To create this pool, we first addressed the lack of established benchmarks by manually writing five seed programs using the Qiskit library. To further diversify our seeds, we then use ChatGPT (see prompt in~\autoref{fig:quantum-prompt}) to generate an additional 15 programs, ensuring that they use a standard set of gates supported across all targeted libraries.

\begin{figure}[tb]
\centering
\begin{lstlisting}[title={Prompt of Generating New Seed Programs}]
**Context**: Existing files [attached].

**Model Role**: Expert quantum computing researcher and programmer

**Task Description**: 
Based on the given quantum program, generate new variant by fulfilling the following requirements:

1. Different from the attached file. The larger difference the better.
2. Try to use conditional statements such as while, if, but not limited to them.
3. Please skip the execution code. Only modifications in the quantum_program function are allowed.
4. The size of the generated program should be similar to the original one.
5. The newly generated code should differ from all previously generated ones.
6. Please remove all comments from the program.
7. Please balance the amount of classical and quantum variables.
8. The quantum gate you can use includes "rx, cz, swap, rz, x, crx, cp, z, crz, h, cx, ry, cry, cswap, ccx, s, t".
\end{lstlisting}
\caption{Prompt for generating new seed programs by using ChatGPT}
\label{fig:quantum-prompt}
\end{figure}

From each seed program, QSPE generates a large family of syntactically diverse but semantically related variants. The QSPE algorithm first parses the seed program into an Abstract Syntax Tree (AST) and identifies all enumeration points, including classical variables and quantum parameters. Then a series of enumeration strategies is applied to these points to produce the final set of variants. The detailed mechanics of the QSPE algorithm are presented in~\autoref{sec:4 qspe}.

\subsection{Step 2: Adding Optimization}

Like other classical platforms, such as C and Java, quantum libraries provide optimization configurations during program compilation. To better exercise our approach and detect more bugs, we add program variants with different optimization levels. That is, for a quantum program variant, we generate three additional variants at optimization levels 1, 2, and 3. The optimization levels used in our approach are summarized in~\autoref{tbl:opt}. We collect optimization strategies from their official documentation of Qiskit\footnote{\url{https://docs.quantum.ibm.com/guides/set-optimization}}, Cirq\footnote{\url{https://quantumai.google/cirq/tutorials/google/spin_echoes}}, and Pytket\footnote{\url{https://docs.quantinuum.com/tket/api-docs/_modules/pytket/backends/backend.html\#Backend.default_compilation_pass}}. Note that Cirq does not provide built-in optimization levels. Therefore, based on the official documentation, we collect the optimizers and manually categorize them into three levels. For Qiskit and Pytket, the optimization levels are already set by the developers, so we directly adopt them in our approach.

\begin{table}[tb]
  \centering
  \caption{Optimization levels adopted in our approach.}
  \label{tbl:opt}
  \begin{tabular}{cccP{4.5cm}}
    \toprule
    Library & Built-in & Level & Optimzation  \\
    \midrule
    Qiskit & Yes & 1 & Layout optimization; InverseCancellation; 1Q gate optimization \\
     &  & 2 & Level 1 + deeper layout optimization; CommutativeCancellation \\
     &  & 3 & Level 2 + deepest layout optimization; Resynthesis of two-qubit blocks; Unitary-breaking passes \\
    \midrule
    Cirq & No & 1 & Drop empty moments; Defer measurement\\
     &  & 2 & Expand composite; Merge single qubit gates to phxz; Stratified circuit\\
     &  & 3 & Eject phased paulis; Drop negligible operations; Eject z; Optimize for target gateset\\
    \midrule
    Pytket & Yes & 1 & Additionally performs some light optimizations \\
     &  & 2 & Add more computationally intensive optimizations that should give the best results from execution \\
    \bottomrule
  \end{tabular}
\end{table}

\subsection{Step 3: Multi-library Program Variants Generation}

In Step 3, we generate program variants in multiple libraries. To simplify implementation, the programs after QSPE enumeration and optimization are written using the Qiskit library. In this step, these programs are translated into quantum programs written by the Cirq and Pytket libraries. Intuitively, we make Qiskit programs $\rightarrow$ Cirq programs, and Qiskit programs $\rightarrow$ Pytket programs in this step.

The biggest challenge in this step is gate compatibility, which refers to the fact that a quantum gate may not be supported by multiple libraries. This is due to the different implementations of the libraries.
After our investigation, we found that Qiskit supports the most extensive range of quantum gates. However, some complex gates supported by Qiskit are not supported in Cirq and Pytket, such as the "cxx" gate and the "rzz" gate. Although we can compose simple gates to synthesize these complex gates, considering that this would introduce significant risks to the correctness of our approach, we have to give up these unsupported gates. Finally, the following gates are used in our approach: ``rx, cz, swap, rz, x, crx, cp, z, crz, h, cx, ry, cry, cswap, ccx, s, t''.

However, it does not mean that the above gates can be perfectly adapted to the Cirq and Pytket libraries. For example, the phase gate ``p gate'' in Qiskit is not supported in Cirq. To address this, we use the ``ZPowGate'' instead, as this gate can replace the ``p gate'' according to Cirq documentation. The ``p gate'' in Qiskit is also not supported by Pytket. We use the "U1 gate" as it has the same functionality as the one in Qiskit. In summary, we make every effort to add more gates for the three libraries.

Assume $N$ programs are generated by QSPE enumeration. Let $\mathcal{P} = \{P_1, P_2, \ldots, P_N\}$ be the set of all generated programs. After step 2, each program $P_i$ generates four variants corresponding to optimization levels 0, 1, 2, and 3, denoted $P_i^0$, $P_i^1$, $P_i^2$, and $P_i^3$, respectively. After step 3, each optimized program $P_i^j$ (where $j \in \{0, 1, 2, 3\}$) further spawns three additional variants: $P_i^{j,Q}$, $P_i^{j,C}$ and $P_i^{j,P}$, where $Q$, $C$ and $P$ denote the three quantum libraries. In total, $12N$ variants are generated before execution.

\begin{table}[tb]
\small
  \centering
  \caption{Testing metrics in our differential testing. $SV(P^{0,Q})$ denotes the statevector of a Qiskit program compiled with optimization level 0 (i.e., without optimization); similar notations apply to other libraries and optimization levels. $Vdot$ denotes the dot product of the two vectors.}
  \label{tbl:criteria}
  \begin{tabular}{cccP{4cm}}
    \toprule
    Rule \# & Program 1 & Program 2 & Metric  \\
    \midrule
    1 & $SV(P^{0,Q})$ & $SV(P^{0,C})$ & $Vdot(SV(P^{0,Q}), SV(P^{0,C})) = 1$ \\
    2 & $SV(P^{0,Q})$ & $SV(P^{0,P})$ & $Vdot(SV(P^{0,Q}), SV(P^{0,P})) = 1$  \\
    \midrule
    3 & $SV(P^{0,Q})$ & $SV(P^{1,Q})$ & $Vdot(SV(P^{0,Q}), SV(P^{1,Q})) = 1$  \\
    4 & $SV(P^{1,Q})$ & $SV(P^{2,Q})$ & $Vdot(SV(P^{1,Q}), SV(P^{2,Q})) = 1$  \\
    5 & $SV(P^{2,Q})$ & $SV(P^{3,Q})$ & $Vdot(SV(P^{2,Q}), SV(P^{3,Q})) = 1$  \\
    \midrule
    6 & $SV(P^{0,C})$ & $SV(P^{1,C})$ & $Vdot(SV(P^{0,C}), SV(P^{1,C})) = 1$  \\
    7 & $SV(P^{1,C})$ & $SV(P^{2,C})$ & $Vdot(SV(P^{1,C}), SV(P^{2,C})) = 1$  \\
    8 & $SV(P^{2,C})$ & $SV(P^{3,C})$ & $Vdot(SV(P^{2,C}), SV(P^{3,C})) = 1$  \\
    \midrule
    9 & $SV(P^{0,P})$ & $SV(P^{1,P})$ & $Vdot(SV(P^{0,P}), SV(P^{1,P})) = 1$  \\
    10 & $SV(P^{1,P})$ & $SV(P^{2,P})$& $Vdot(SV(P^{1,P}), SV(P^{2,P})) = 1$  \\
    \bottomrule
  \end{tabular}
\end{table}

\subsection{Step 4: Execution and Result Analysis}

In Step 4, we run all the programs and analyze the output results based on the differential comparison rules and metrics, which will be discussed below.

\subsubsection{Validation Method.} 
\begin{figure}[!tb]
\centering
\includegraphics[width=0.78\textwidth]{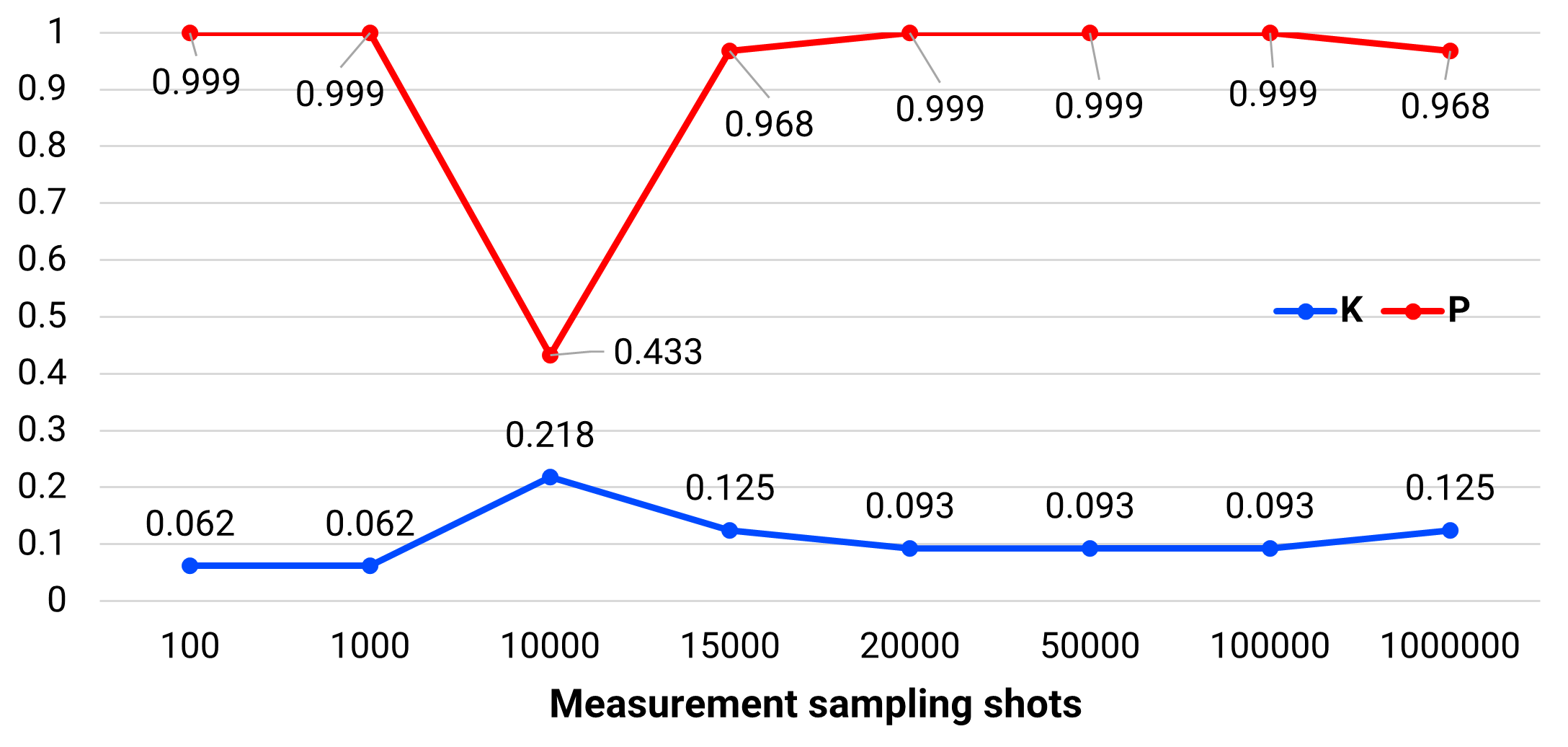}
\caption{The K and P values with the increasing measurement sampling shots.}
\label{fig:measure}
\end{figure}

In the previous approach, researchers used the measurement-based validation method, which involves calculating the distance between distributions of measurement sampling results to verify program behaviors. 
However, we have identified that the measurement-based validation method often yields incorrect results. Due to the non-deterministic nature of quantum programs, increasing the sampling shots allows the measurement results to approximate the ground truth distribution. However, it is challenging to determine whether the approximation is sufficient or accurately represents the actual distribution. 
As illustrated in~\autoref{fig:measure}, we use the K-S test~\cite{kolmogorov1933sulla} to compare two distributions, as introduced in QDiff. We collect the K and P values, which are the two essential factors in the K-S test. In QDiff, if the K value of two programs is greater than an empirical threshold $t$, such as 0.15, it indicates that abnormal behaviors are present in the programs. However, as we collect K and P values for different sampling shots, we find that the values vary a lot, as shown in~\autoref{fig:measure}.

\begin{figure}[!tb]
\centering
\includegraphics[width=0.78\textwidth]{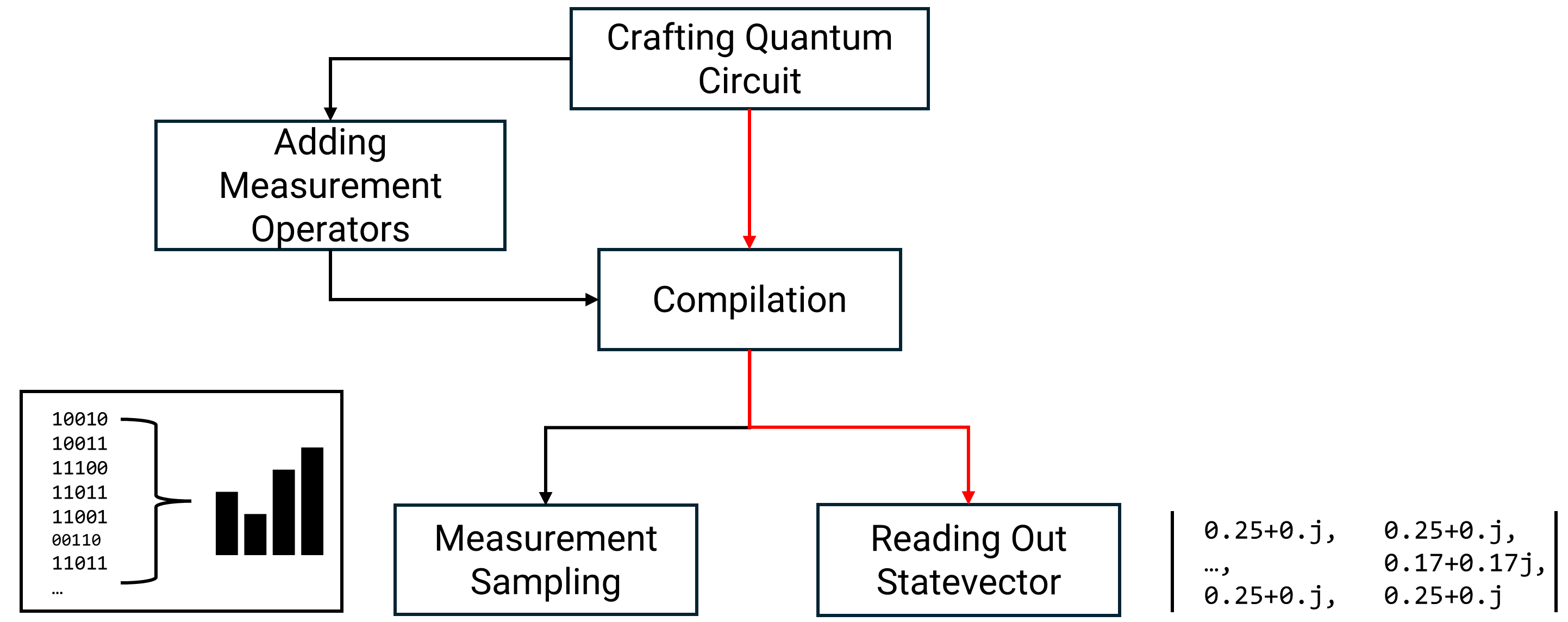}
\caption{The comparison of the process of getting statevector (in red arrow) and collecting samplings to a distribution (in black arrow).}
\label{fig:process}
\end{figure}

To eliminate validation instability, we propose using statevector-based validation in practice. Unlike measurement-based validation, the statevector-based method does not require the addition of measurement operations to the circuit, as illustrated in~\autoref{fig:process}. Statevector-based validation shows three key advantages: (1) Stability. Statevector comparison provides deterministic results by directly comparing quantum state amplitudes, eliminating the statistical uncertainty inherent in measurement sampling that relies on probability distributions varying across runs; (2) Efficiency. Statevector extraction requires only a single simulation run to obtain complete quantum state information, while measurement-based validation typically requires thousands of sampling runs (e.g., 10,000 shots) to construct probability distributions; and (3) Interpretability. Measurement operations cause quantum state collapse, leading to irreversible loss of quantum information and phase relationships, whereas statevectors preserve the complete quantum state, including amplitude and phase information, enabling result reproduction and facilitating downstream application.

\subsubsection{Variant Comparison.} We propose testing metrics for differential testing across multiple optimization levels and quantum computing libraries. As shown in~\autoref{tbl:criteria}, we comprehensively compare statevectors from programs across three libraries (Qiskit, Cirq, and Pytket) and four optimization levels (0, 1, 2, 3) by calculating the dot product of statevectors. We design comparison strategies based on optimization level considerations. 

For unoptimized programs (i.e., optimization level 0), we conduct cross-library comparisons since these programs should produce identical results regardless of the underlying library implementation. For example, Rule 1 compares the output between the Qiskit program $P_{i}^{0,Q}$ and the Cirq program $P_{i}^{0,C}$, while Rule 2 compares the Qiskit program $P_{i}^{0,Q}$ with the Pytket program $P_{i}^{0,P}$.

For optimized programs (i.e., optimization levels 1, 2, and 3), cross-library comparisons are not included due to implementation differences in optimization strategies across libraries that may introduce systematic bias. Instead, we perform intra-library comparisons between different optimization levels. For example, we compare the Qiskit program with optimization level 1 against the unoptimized version: $P_{i}^{1,Q}$ vs. $P_{i}^{0,Q}$. This approach ensures that any differences observed are due to optimization transformations rather than library-specific implementation variations.


%% file: sec/4-qspe.tex
\section{Quantum Skeletal Program Enumeration}
\label{sec:4 qspe}

In this section, we introduce the algorithm for our quantum skeletal program enumeration. Specifically, in~\autoref{subsec:4-1}, we briefly introduce the classical approach of SPE; in~\autoref{subsec:4-2}, we extend the classical approach to quantum variables and discuss the differences compared with the classical approach; in~\autoref{subsec:4-3}, we summarize our implementation as an algorithm to provide a straightforward representation of our approach.

\subsection{$\alpha$-equivalent Program}
\label{subsec:4-1}

\subsubsection{$\alpha$-equivalent Program} 

Given a program $P$, the program skeleton is a structural representation that replaces variable names with placeholders while preserving control-flow information. A naive approach to generating new programs from the skeleton is arbitrarily filling these placeholders with variables from the original program. For example, given the program skeleton in~\autoref{fig:spe example}, which contains three variables $\{a, b, c\}$ and seven holes, the total number of possible variants is $3^7 = 2,187$. 

However, the naive method suffers from two critical limitations: 1) exponential complexity that makes exhaustive enumeration computationally infeasible for larger programs, and 2) semantic redundancy that most enumerated programs represent trivial modifications that differ only through $\alpha$-conversion (variable renaming). Such redundancy fails to effectively explore unseen program behaviors. Given a program $P$, if another program $P'$ is obtained from $P$ via $\alpha$-conversion, then $P$ and $P'$ are considered $\alpha$-equivalent. In~\autoref{fig:spe example}, considering the classical variable holes $\square$, program $P$ uses the variable array $\{a, b, a, c, b, c, a\}$, while $P_1$ uses $\{b, a, b, c, a, c, b\}$. The only difference between the two arrays is the swapping of variables $a$ and $b$. Since the renaming does not change the program’s semantics, $P$ and $P_1$ are $\alpha$-equivalent. Such equivalent programs typically share the same control-flow structure and produce identical execution results, offering limited value for testing.

\input{figure/alpha-eq}

\subsubsection{Partitions with Scopes}

To improve the efficiency of enumeration, $\alpha$-equivalent programs must be eliminated. Zhang et al.~\cite{zhang2017skeletal} propose a partition-based enumeration approach that avoids generating $\alpha$-equivalent variants. Instead of directly filling holes with variables, their method focuses on enumerating partitions of holes first and then assigning variables to each partition.

Given a program $P$ with $n$ variables $V = \{v_{1}, v_{2}, \ldots, v_{n}\}$ and $k$ holes $H = \{h_{1}, h_{2}, \ldots, h_{k}\}$, the number of distinct partitions of $H$ into $n$ non-empty subsets corresponds to the Stirling number of the second kind, denoted as $\stirlingII{k}{n}$~\cite{donald1999art}. Each partition defines a mapping from hole subsets to variables, ensuring that variable renamings do not result in redundant $\alpha$-equivalent programs.

For example, in the program $P$ shown in~\autoref{fig:spe example}, ignoring quantum variables such as \textit{qc}, there are two classical variables $V = \{a, b\}$ in the global scope and seven holes $H = \{h_{1}, h_{2}, \ldots, h_{7}\}$. The number of ways to partition the holes into two subsets is $\stirlingII{7}{2} = 63$. Each partition, such as $\{\{h_{1}, h_{2}, \ldots, h_{7}\}, \{\}\}$, $\{\{h_{1}, h_{2}, \ldots, h_{6}\}, \{h_{7}\}\}$, defines a unique variable-to-hole assignment, resulting in one distinct program variant. In total, this yields 63 $\alpha$-inequivalent enumeration results under this skeletal program enumeration (SPE) strategy.

Another important factor in enumeration is scope constraints, which restrict where variables can be placed. A variable can only be assigned to holes within its declaring scope or in nested subscopes. To ensure scope correctness and avoid generating invalid code, SPE performs enumeration in a scope-aware manner: it counts variables and holes within each scope separately, and computes the final enumeration results by taking the Cartesian product over all local scope combinations.

Given a program $P$ with a set of scopes $S = \{S_g, S_1, S_2, \ldots, S_t\}$, where $S_g$ denotes the global scope present in every program, the complete enumeration set is defined as $F = S_g \times S_1 \times S_2 \times \ldots \times S_t$. For example, in the program shown in~\autoref{fig:spe example}, variable \texttt{c} is declared within the inner scope $S_1$ under the \texttt{if} branch, and thus can only be assigned to holes within positions ${4,5,6,7}$. In contrast, variables \texttt{a} and \texttt{b}, declared in the global scope $S_g$, can be assigned to any hole in the program. One of the final enumeration results can be the Cartesian product of all scope-respecting partitions: $\{\{1,5\}, \{2,3\}, \{4,6,7\}\}$, as shown in the program $P_2$.

\subsection{Classical Variable Enumeration Meet Quantum Variable}
\label{subsec:4-2}

The Quantum Skeletal Program Enumeration (QSPE) algorithm integrates classical SPE techniques for enumerating traditional variables with quantum-specific strategies adapted to quantum circuit parameters and qubit assignments. This hybrid approach addresses the limitations of classical SPE, which cannot adequately handle quantum-specific components, that often constitute more than 50\% of a quantum program.

Quantum components introduce two additional enumeration dimensions: rotation angles and target qubits. Unlike variable holes in the classical SPE approach, rotation angles and target qubits are parameters that can be adjusted. We observe that modifying the parameters can lead to significant changes in the behavior of the quantum program. Therefore, in our approach, we treat these two parameters as targets for variation. Based on this, we define the quantum-$\alpha$-equivalence as the following:


\begin{definition}[Quantum-$\alpha$-Equivalence]
Two quantum programs are quantum-$\alpha$-equivalent when they execute the same sequence of gates with identical parameters, except that the qubit indices are permuted (e.g., swapping qubit $q_1$ and $q_2$ yields an equivalent program).
\end{definition}

Specifically, rotation angles apply to parameterized quantum gates such as $R_x(\theta)$, $R_y(\theta)$, and $P(\theta)$, where the angle $\theta$ can significantly influence the program behavior. The target qubits specify the qubits on which a gate operates. For classical variables within the quantum program, QSPE directly adopts the original SPE method. For quantum variables, QSPE extends the enumeration to explore both gate parameters and qubit assignment.

To handle rotation angle parameters, we adopt a random sampling strategy. Specifically, each angle $\theta$ is sampled from the interval $[0, 2\pi)$. In previous studies, the commonly used angle parameter is from $\{0, \frac{\pi}{4}, \frac{\pi}{2}, \pi, \frac{3\pi}{2}\}$. To enhance enumeration diversity, we randomly generate angle values within the valid range. Given a program $P$ with $k$ parameterized rotation gates, we generate a parameter set $\Theta = \{\theta_{1}, \theta_{2}, \ldots, \theta_{k}\}$, where each $\theta_i$ corresponds to the angle applied to the $i$-th rotation gate.

For qubit assignment enumeration, to make a diverse enumeration output, we collect the target qubit placeholders as holes in quantum programs. Given a program $P$ with $q$ target qubit parameters, we generate a parameter set $Q = {Q_1, Q_2, \ldots, Q_q}$ that denotes the set of all qubit assignment holes in the program. Each $Q_i$ specifies a position that requires a qubit assignment. We then shuffled the order of the holes and assigned qubits to the holes. Note that there are two types of qubit assignments: target qubits and control qubits. We carefully assign qubits to holes to ensure correctness and preserve program validity.

The enumeration of the qubit must follow the two constraints to ensure the validity of the program.

\begin{enumerate}
    \item Qubit range restriction: Each qubit assignment must refer to a valid index within the allocated quantum register. If a gate references a qubit outside the defined range, the compiler will raise an exception and terminate execution. 
    \item Target-control conflict: The same qubit cannot be simultaneously assigned as both the target and control in a multi-qubit gate. Such conflicts result in compiler errors, rendering the program invalid for testing.
\end{enumerate}

Classical $\alpha$-equivalence can be extended to quantum programs, but quantum variables require additional definitions of equivalence. For rotation angle parameters, equivalence requires that the angles be numerically equal. For qubit assignments, equivalence appears when programs differ only in the swapping of qubits (e.g., swapping the first and second qubits) while preserving functional behavior. This is analogous to classical variable renaming, but applied within the quantum register space.

\subsection{Quantum-Classical Hybrid Enumeration Algorithm}
\label{subsec:4-3}

The core algorithm integrates the classical SPE methodology with quantum-specific enumeration techniques through a three-phase approach, as shown in~\autoref{alg:qspe}. The first phase applies classical SPE to enumerate all non-$\alpha$-equivalent classical variable assignments, producing the set of classical program skeletons. The second phase systematically enumerates the quantum parameters for each quantum variable. The third phase combines the two approaches and generates the programs.

\begin{algorithm}
\caption{Quantum Skeletal Program Enumeration}
\label{alg:qspe}
\begin{algorithmic}[1]
\REQUIRE Program $P$
\ENSURE Non-$\alpha$-equivalent program set $\mathcal{T}$ \\
\textcolor{teal}{// Phase 1: Classical Variable Enumeration}\\
\STATE Enumerations of classical holes $\mathcal{E}_{classical} \gets \emptyset$
\STATE variables scopes $S = \{S_g, S_1, \ldots, S_t\}$, max qubits $N_{qubit}$ $\gets ASTAnalysis(P)$
\FOR{$S_i \in S_t$}
    \STATE $V \gets$ variables in scope $S_i$
    \STATE $H \gets$ holes in scope $S_i$
    \STATE $\Omega \gets$ all partitions with $\stirlingII{|V|}{|H|}$ 
    \STATE $\mathcal{E}_i \gets$ \{enumerations from $\Omega$ respecting scope $S_i\}$
\ENDFOR
\STATE $\mathcal{O}_{classical} \gets \mathcal{E}_g \times \mathcal{E}_1 \times \cdots \times \mathcal{E}_t$ \\
\textcolor{teal}{// Phase 2: Quantum Variable Enumeration} \\
\FOR{$S_i \in S_t$}
    \STATE $\theta \sim [0, 2\pi)$
    \STATE $H \gets$ quantum holes in scope $S_i$
    \IF{H is for qubit assignment}
        \REPEAT  
            \STATE $Q \gets$ selected from $[0, N_{qubit})$
        \UNTIL{$\textsc{notQuantumAlphaEQ}(H, Q, S_i)$} \\
            $\mathcal{E}_i \gets Q$
    \ELSIF{H is for rotation angle}
        \STATE $\mathcal{E}_i \gets \theta$
    \ENDIF
\ENDFOR
\STATE $\mathcal{O}_{quantum} \gets \mathcal{E}_1 \times \cdots \times \mathcal{E}_t$ \\
\textcolor{teal}{// Phase 3: Program Generation} \\
\STATE $\mathcal{T} \gets \emptyset$
\WHILE{not empty $\mathcal{O}_{classical}$ or $\mathcal{O}_{quantum}$}
    \STATE $\mathcal{T} \gets$ $FillHoles(\mathcal{O}_{classical}, \mathcal{O}_{quantum}, P) $
\ENDWHILE
\RETURN $\mathcal{T}$

\end{algorithmic}
\end{algorithm}

\subsubsection{Phase 1: Classical Variable Enumeration}
The algorithm starts by identifying all scopes in the program where classical variables can be assigned, including the global scope $S_g$ and each individual local scope $S_1, \ldots, S_t$. For each scope $S_i$, the algorithm extracts the set of declared variables $V$ and the corresponding set of holes $H$. Then we construct a mapping from variables to holes in a way that follows scope constraints and avoids illegal semantics. 

To systematically explore the space of such mappings, the algorithm computes all possible partitions of the variable set $V$ into $|H|$ non-empty subsets. The number of such partitions is given by the Stirling number of the second kind, $\stirlingII{|V|}{|H|}$. Each partition represents a different strategy to fill the holes in that scope. These local enumerations $\mathcal{E}_i$ are collected for the scope $S_i$, and combined using a Cartesian product to produce the entire enumeration space $\mathcal{O}{classical}$ for classical variables.

\subsubsection{Phase 2: Quantum Variable Enumeration}
After enumerating the classical variables, the algorithm turns to filling in quantum-specific holes, which correspond to qubit assignments (which qubit a gate should act on) and rotation angles (e.g., for parameterized gates like $R_y(\theta)$).

For qubit assignments, the algorithm randomly selects a qubit index $Q$ from the available range $[0, N_{qubit})$, where $N_{qubit}$ is the number of qubits registered in the quantum program. However, not all assignments are semantically valid. For example, assigning both control and target qubits of a two-qubit gate to the same qubit violates the semantics of the quantum gate. Therefore, we use a \texttt{notQuantumAlphaEQ} function to check each candidate assignment against the quantum-$\alpha$-equivalence definition and the known constraints. Here, \texttt{notQuantumAlphaEQ} returns true when the new statement, formed by filling hole $H$ in scope $S_{i}$ with qubits $Q$, is not quantum-$\alpha$-equivalent to the original statement, and false otherwise. For rotation parameters, a random sampling strategy is used. A representative angle $\theta$ is sampled from the interval $[0, 2\pi)$ to span the relevant parameter space. The enumerations across scopes yield the quantum instantiation space $\mathcal{O}_{quantum}$ for quantum variables.

\subsubsection{Phase 3: Program Generation}
In the final phase, QSPE generates full candidate programs by combining the results of classical and quantum enumeration. We use the product of $\mathcal{O}{classical} \times \mathcal{O}{quantum}$ to generate new combinations of variables to fill the holes in the skeletal program $P$. These filled programs are stored in the result set $\mathcal{T}$. 

%% file: figure/alpha-eq.tex
\begin{figure}[tb]
\centering
\lstset{
    language=Python,                
    basicstyle=\footnotesize\ttfamily,
    keywordstyle=\bfseries\color{blue}, 
    stringstyle=\color{red!70!black},   
    commentstyle=\color{green!50!black},
    identifierstyle=\color{black},      
    backgroundcolor=\color{white},      
    frame=none,                         
    xleftmargin=.5\parindent,
    xrightmargin=.5\parindent,
    aboveskip=2pt,
    belowskip=2pt,
    columns=fullflexible,
    keepspaces=true,
    escapeinside={|}{|},                
    showstringspaces=false              
}

\begin{subfigure}{0.22\textwidth}
\centering
\begin{lstlisting}
def main():
    a = 1
    b = 0
    qc.t(0)
    if a:
        qc.rx(0.123, 1)
        c = 3
        b = c + a
\end{lstlisting}
\caption{Program $P$}
\label{fig:6-1}
\end{subfigure}
\hfill
\begin{subfigure}{0.22\textwidth}
\centering
\begin{lstlisting}
def main():
    |$\square$| = 1
    |$\square$| = 0
    qc.t(|$\bigcirc$|)
    if |$\square$|:
        qc.rx(|$\triangle$|, |$\bigcirc$|)
        |$\square$| = 3
        |$\square$| = |$\square$| + |$\square$|
\end{lstlisting}
\caption{Skeleton $\mathbb{P}$}
\label{fig:6-2}
\end{subfigure}
\hfill
\begin{subfigure}{0.22\textwidth}
\centering
\begin{lstlisting}
def main():
    b = 1
    a = 0
    qc.t(0)
    if b:
        qc.rx(0.123, 1)
        c = 3
        a = c + b
\end{lstlisting}
\caption{Program $P_1$}
\label{fig:6-3}
\end{subfigure}
\hfill
\begin{subfigure}{0.22\textwidth}
\centering
\begin{lstlisting}
def main():
    a = 1
    b = 0
    qc.t(2)
    if b:
        qc.rx(0.3, 0)
        c = 3
        a = c + c
\end{lstlisting}
\caption{Program $P_2$}
\label{fig:6-4}
\end{subfigure}

\caption{An example of a program skeleton $\mathbb{P}$ and generation results $P_1$ and $P_2$. $\square$ denotes the classical variable hole, $\triangle$ denotes the quantum rotation angle hole and $\bigcirc$ denotes the quantum qubit assignment hole.}
\label{fig:spe example}
\end{figure}

%% file: sec/5-exp.tex
\section{Evaluation}
\label{sec:evaluation}

QSPE is implemented in Python in 3,298 lines of code. All experiments are conducted on a machine running Ubuntu 18.04 LTS, equipped with an Intel Xeon E5-2620v4 CPU, 32GB RAM, and a 2TB HDD. 
To evaluate the effectiveness of QSPE, we evaluated QSPE using three mainstream quantum programming libraries: Qiskit version 2.0.0, Cirq version 1.4.1, and Pytket version 2.3.2. Comparisons with previous approaches, such as QDiff and MorphQ, are not included in our experiments because our approach focuses on a very different task that makes the direct comparison unavailable. We aim to answer the following research questions.
\begin{enumerate}[leftmargin=*,label=\textbf{RQ\arabic*.},topsep=0pt,itemsep=0ex]
\item How efficient is QSPE in generating valid programs and reducing redundant enumeration?
\item By adopting statevector-based validation, can QSPE produce fewer false positives?
\item What has QSPE found via differential testing of the three compilers?
\end{enumerate}

RQ1 assesses the efficiency of QSPE by measuring both the number of valid programs it generates and the time saved by avoiding $\alpha$-equivalent variants. For RQ2, we analyze the failure cases found in the experiments to demonstrate the effectiveness of statevector-based validation in capturing abnormal behavior. In RQ3, we summarize the miscompilations detected during differential testing and report the confirmed issues to the official repositories for developer review. We also include a brief discussion of the root causes and consequences of these miscompilations. Due to space limitations, this section focuses on the summarized results. Full experimental configurations and detailed results are available on our website at: \url{https://sites.google.com/view/qspe/}.

\subsection{RQ1: enumeration reduction}

Due to the lack of a standard quantum program benchmark and the difficulty of estimating the outputs of arbitrary quantum programs, we manually wrote five representative seed programs. Furthermore, we generated 15 more seed programs using ChatGPT (see~\autoref{sec:3 approach}), resulting in a total of 20 seed quantum programs for the evaluation of QSPE. These programs range from 35 to 56 lines of code. Programs with more than 56 lines are not included because large programs tend to produce dead loops in executions.

\begin{figure}[tb]
  \centering
  \begin{subfigure}[t]{0.48\textwidth}
    \centering
    \includegraphics[width=\linewidth]{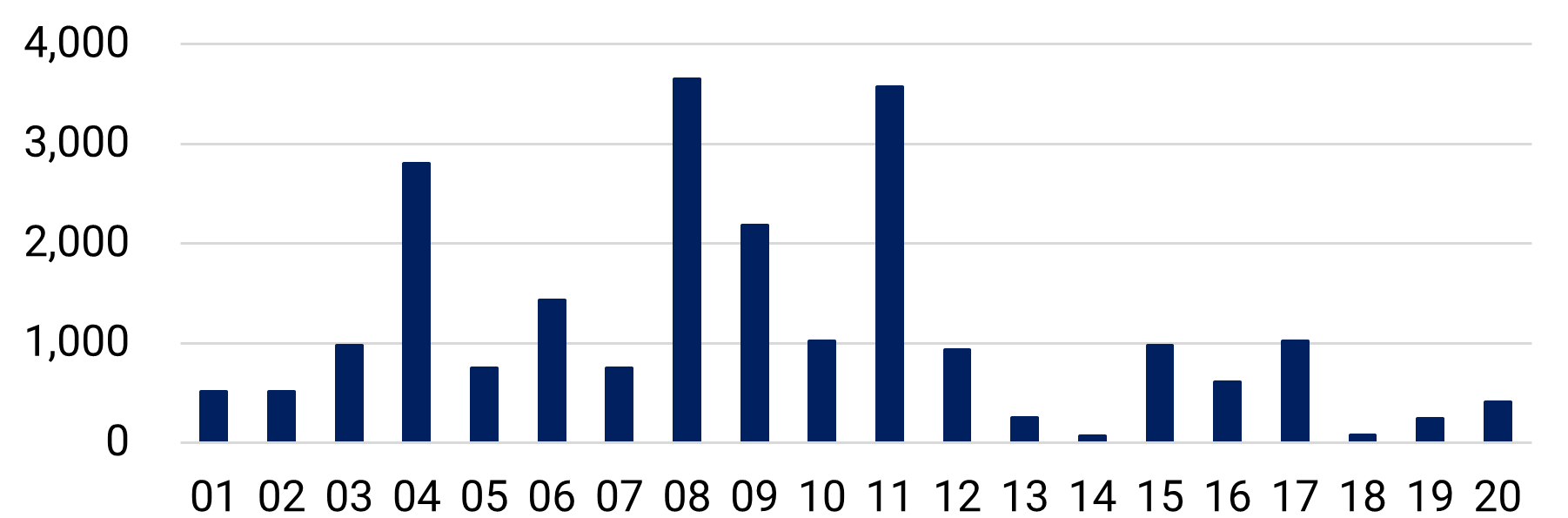}
    \caption{Distribution of the numbers of generated variants by applying the QSPE approach to our 20 seed quantum programs.}
    \label{fig:rq1-1}
  \end{subfigure}
  \hfill
  \begin{subfigure}[t]{0.48\textwidth}
    \centering
    \includegraphics[width=\linewidth]{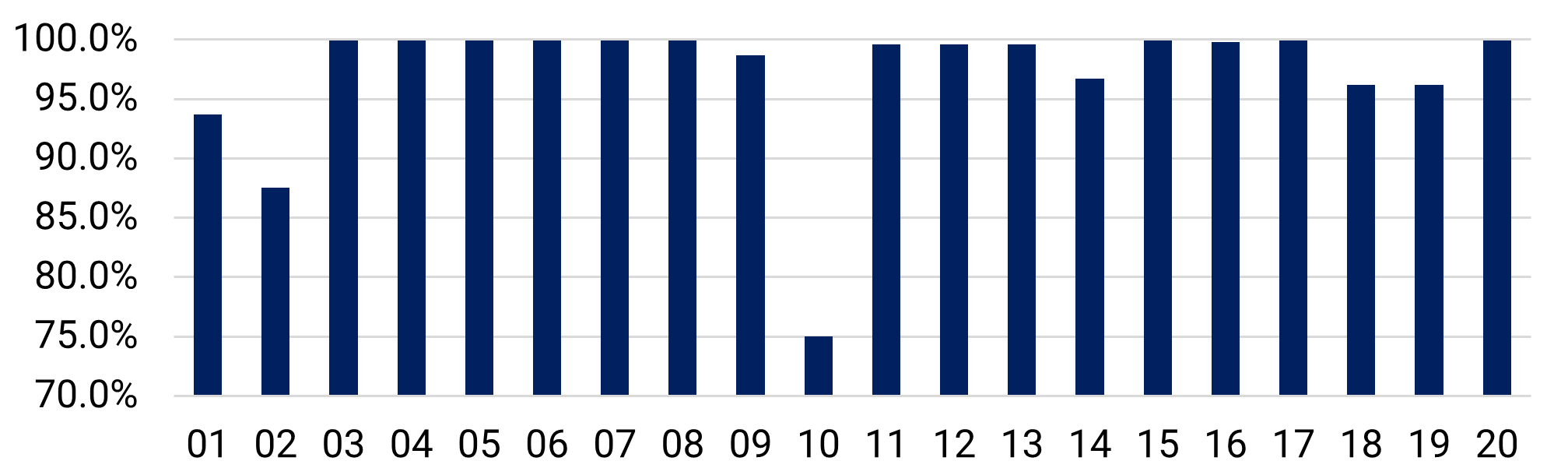}
    \caption{Distribution of the ratios of the reduced variants. For example, the first bar shows 93.7\% of the variants are reduced compared with the na\"{i}ve enumeration method.}
    \label{fig:rq1-2}
  \end{subfigure}
  \caption{Distributions of the variants reduction. In both subfigures, the $x$-axis denotes the 20 seed quantum programs adopted in our experiments.}
  \label{fig:rq1-both}
\end{figure}

\begin{figure}[!tb]
\centering
\includegraphics[width=0.68\textwidth]{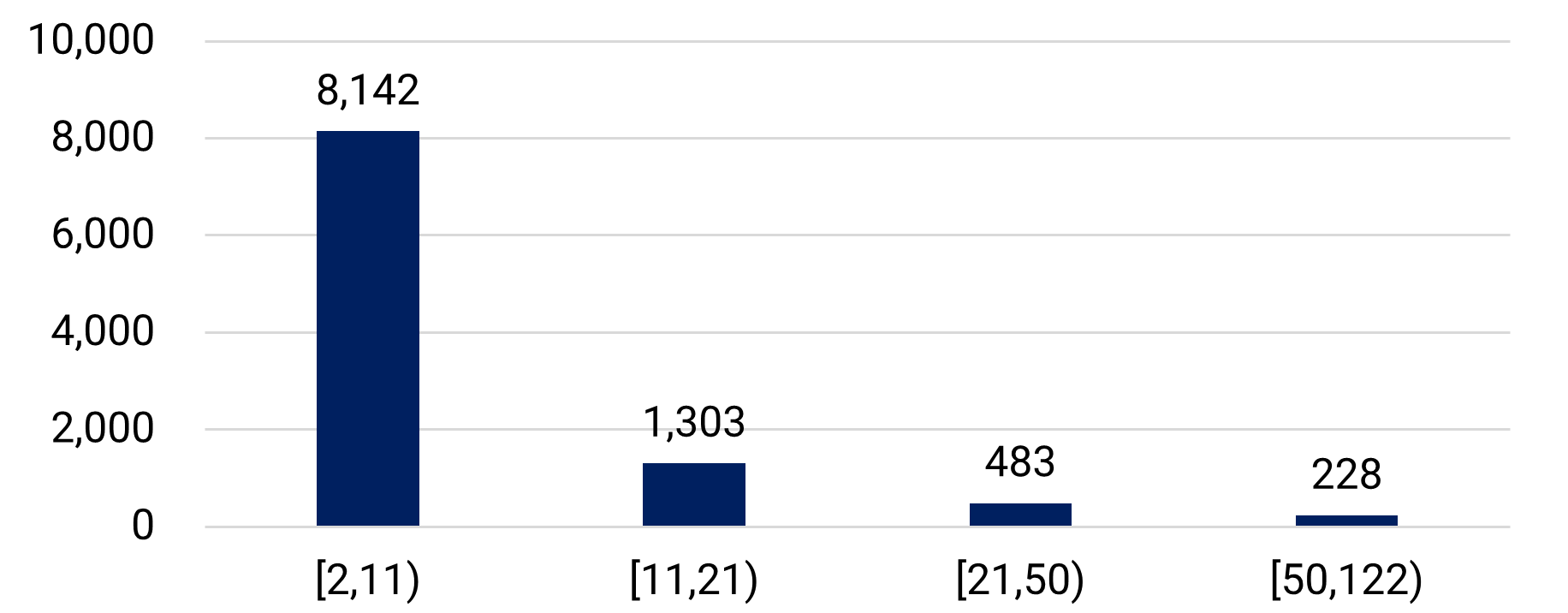}
\caption{The distributions of the number of gates of the circuit of the enumeration variants. 
}
\label{fig:rq1-3}
\end{figure}

\autoref{fig:rq1-both} summarizes the number of variants generated from the 20 seed programs using QSPE under the Qiskit backend without optimization. To simplify the analysis, we exclude the results for Cirq and Pytket in this part. \autoref{fig:rq1-1} illustrates the number of variants produced for each seed program. The counts range from 70 variants (seed 14) to 3,650 variants (seed 08). 
\autoref{fig:rq1-2} shows the reduction in the number of variants compared to the naive enumeration approach. For 18 out of 20 seed programs, QSPE reduces more than 90\% of variants. Even for seed 10, which has the lowest reduction rate, over 70\% of variants are eliminated. Notably, seed programs with fewer variants tend to have lower reduction rates, such as seeds 18 and 19, each producing fewer than 250 variants. This indicates a trade-off between variant quantity and reduction rates.

We further examine the diversity of circuit sizes covered by QSPE-generated variants. \autoref{fig:rq1-3} shows the distribution of circuit sizes, where the $x$-axis represents gate count intervals. For example, the first bar indicates that 8,142 variants are circuits with 2–10 gates. Notably, while most generated circuits are small (2–11 gates), QSPE is capable of producing larger circuits with up to 122 gates. 

\begin{tcolorbox}[size=title,rightrule=1mm, leftrule=1mm, toprule=0mm, bottomrule=0mm, arc=0pt,colback=gray!5,colframe=myblue,breakable]
{ \textbf{Answer to RQ1: } 
In terms of reducing enumeration, the QSPE approach generates 70 to 3,650 variants, reducing more than 90\% of the variants compared to the naive approach. We further investigate the size of the generated circuit that QSPE can cover. The result shows that QSPE is capable of generating circuits with gate counts ranging from 2 to 122. 
}
\end{tcolorbox}

\subsection{RQ2: output validation method}

While QSPE aims to generate valid test programs, an improper validation method can make it fail to effectively expose program differences, resulting in misleading reports. Previous approaches, such as QDiff~\cite{wang2021qdiff} and MorphQ~\cite{paltenghi2023morphq}, adopt measurement-based validation to verify the output of the quantum program. This method relies on sampling from the underlying probability distribution. However, due to the inherently probabilistic nature of quantum computing, measurement-based validation only approximates the true distribution and is sensitive to the number of shots, resulting in fluctuations in the observed results.

We propose to use statevector-based validation in testing. Unlike measurement-based methods that rely on statistical distribution comparisons, statevectors eliminate statistical uncertainties by directly representing the states of the quantum program without probabilistic sampling. The statevectors can be read out after compilation, as shown in~\autoref{fig:process}. Considering that the statevector is prone to being affected by the global phase, after discussion with Qiskit developers, we propose to use the dot product calculation of the statevectors (see~\autoref{sec:3 approach}) for comparing programs' behaviors.

\begin{figure}[!tb]
\centering
\includegraphics[width=0.8\textwidth]{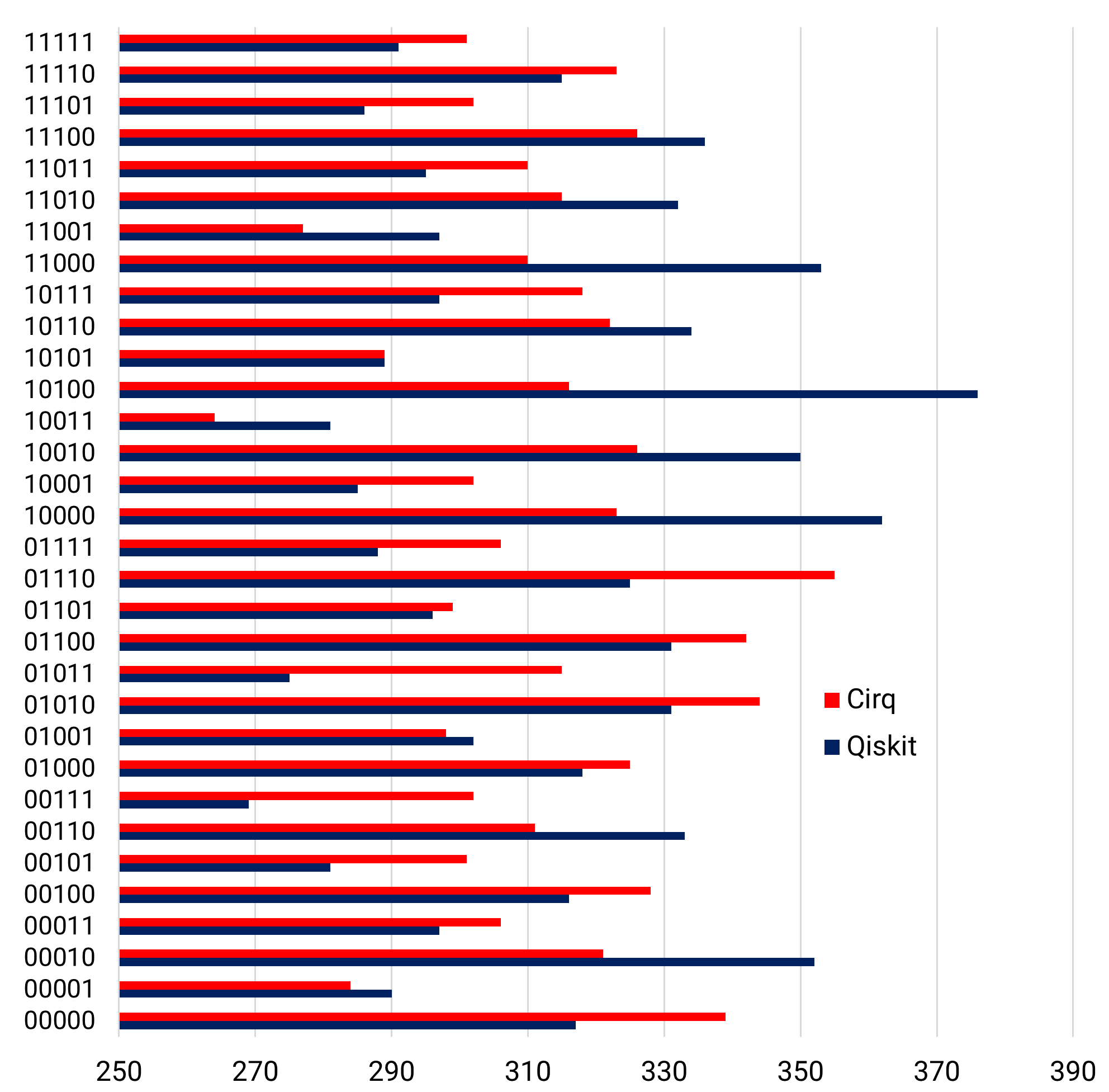}
\caption{The sampling distribution of a quantum program compiled by Cirq (in red color) and Qiskit (in blue color).}
\label{fig:rq2-1}
\end{figure}

\begin{figure}[!tb]
\centering
\includegraphics[width=0.9\textwidth]{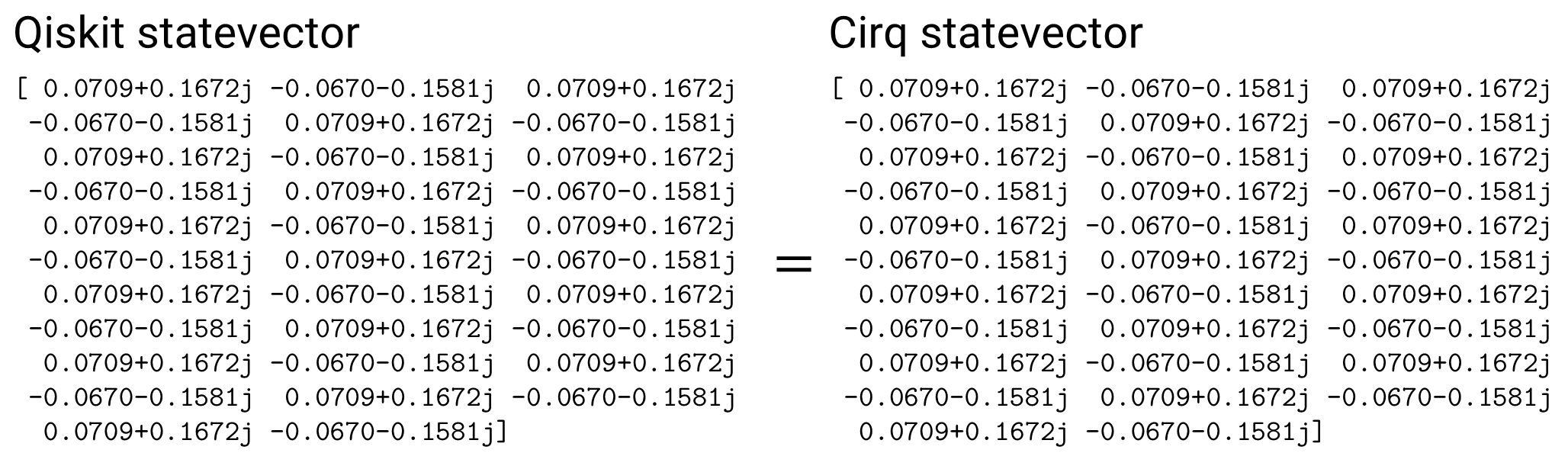}
\caption{The statevectors of the example in~\autoref{fig:rq2-1}}
\label{fig:rq2-2}
\end{figure}

To demonstrate the effectiveness of statevector-based validation, we compare it against measurement-based validation. As an example, we compile the same quantum program using both Qiskit and Cirq, and analyze the resulting output distributions. The sampling results are shown in~\autoref{fig:rq2-1}, where the red bars represent Cirq and the blue bars represent Qiskit. As observed, the two distributions differ significantly, and most outcomes display mismatched frequencies. By using the measurement-based validation method, this example will be considered as a miscompilation between Qiskit and Cirq, due to the divergence between the two distributions.

However, when comparing the statevectors of the two compiled programs, we find them to be identical (see~\autoref{fig:rq2-2}), confirming that the program behavior is semantically equivalent. This indicates that the earlier miscompilation report is a false positive. In our experiments, the measurement-based validation method reports more than 20,000 miscompilations. However, upon further random selection of 100 cases, we find that they are all false alarms. Therefore, we believe that this may be a common case in measurement-based validations.

In fact, the reason why the measurement sampling results vary between the two libraries is that the sampling process introduces inevitable randomness. Although quantum libraries provide configurations to eliminate randomness, such as the sampling seed, at the level of bottom implementation, the Random Number Generation (RNG) system produces different outputs across platforms. In contrast, the statevector-based method captures the full quantum state, including amplitude and phase information, and is not affected by runtime randomness. This illustrates its robustness and precision in comparing the behaviors of quantum programs, making it a reliable alternative to measurement-based validation. By equipping the statevector-based validation to QSPE, we find that the previously reported cases are largely reduced, and hundreds of new bugs are reported. After we further send them to the developers of the quantum libraries, the reports are confirmed as true bugs, as illustrated in our discussion of RQ3.

\begin{tcolorbox}[size=title,rightrule=1mm, leftrule=1mm, toprule=0mm, bottomrule=0mm, arc=0pt,colback=gray!5,colframe=myblue,breakable]
{ \textbf{Answer to RQ2: } 
Previous approaches rely on measurement-based validation to detect miscompilations. However, this method suffers from significant limitations due to the inherent randomness of quantum measurement. To address this issue, we propose using statevector-based validation as an alternative. In experiments, statevector-based validation method successfully reduces false-positive reports. This demonstrates the effectiveness of adopting statevector-based validation in quantum program testing.
}
\end{tcolorbox}

\subsection{RQ3: bug report and discussion}

\begin{table}[tb]
  \centering
  \caption{Miscompilations found in our experiments.}
  \begin{minipage}{0.45\textwidth} 
  \end{minipage}
  \label{tbl:bugs}
  \begin{tabular}{cccc}
    \toprule
    Library & \# of Miscompilation & Violated Rule & Report Status \\
    \midrule
    Cirq & 0 & N.A. & N.A. \\
    Qiskit & 81 & $P^{1,Q} vs P^{2,Q}$ & Approved \\
    Pytket & 4 & $P^{1,P} vs P^{2,P}$ & Approved \\
    Pytket & 623 & $P^{0,P} vs P^{1,P}$ & Approved \\
    \bottomrule
  \end{tabular}
\end{table}

Our comprehensive testing methodology uncovered critical insights into compiler reliability across multiple quantum computing frameworks. By systematically applying our statevector-based differential testing approach, we identified a total of 708 distinct miscompilation cases across two major compilers: Qiskit and Pytket (see~\autoref{tbl:bugs}). Among them, 81 unique miscompilations were observed in Qiskit and 627 unique miscompilations were found in Pytket. For Cirq, we did not identify any bugs before submitting the manuscript, but we will continue our experiments to uncover potential new findings.

These miscompilations were centered on three violated rules: $P^{1,Q}$ vs. $P^{2,Q}$, $P^{1,P}$ vs. $P^{2,P}$ and $P^{0,P}$ vs. $P^{1,P}$. Specifically, the first violation, denoted as $P_Q^1$ vs. $P_Q^2$, captures cases where two quantum circuits are expected to produce identical statevectors but yield divergent outputs after compilation. The second and third violations correspond to inconsistencies introduced by different optimization levels in parameterized circuits. Notably, Pytket exhibited at most 623 miscompilations, which may suggest potential flaws in its parameter handling and optimization mechanisms.

We reported our findings to the development teams of both libraries. Due to the large number of miscompilation cases, it was infeasible to submit all reports individually. Instead, we selected representative cases for submission and inferred whether the remaining instances shared the same root cause based on behavioral similarities. The response from the development teams varied across compilers. The Qiskit team responded promptly, acknowledged all 81 reported miscompilations, and confirmed that the fixes would be incorporated into future development plans. The Pytket team also engaged actively with our reports. Given the larger volume of miscompilations, they started a broader investigation of parameter handling and circuit optimization pipelines. As both categories of violations were acknowledged, so we label the current status of Pytket-related findings as ``Approved''.

The miscompilations of the Qiskit library are due to a missing method in the statevector class in Qiskit. We report this issue to the developers of the official library~\footnote{Link removed for anonymity; will be included in the camera-ready version.}. We consider this a miscompilation because we found that the same quantum programs compiled under both optimization levels 1 and 2 output different statevectors. Meanwhile, the dot product of the two state vectors is 0.45, neither 1.0 (which indicates that the state vectors are in the same direction) nor 0.0 (which means that the state vectors are in the opposite direction), indicating that some unusual program behavior is occurring. Upon further discussion with the Qiskit developers, it was confirmed that the miscompilations resulted from a missing library in the statevector class in Qiskit. 

\begin{tcolorbox}[size=title,rightrule=1mm, leftrule=1mm, toprule=0mm, bottomrule=0mm, arc=0pt,colback=gray!5,colframe=myblue,breakable]
{ \textbf{Answer to RQ3: } 
By using QSPE, we detected 708 miscompilations across quantum libraries, demonstrating the effectiveness of QSPE in identifying bugs in differential testing. After we contacted the developers, 81 miscompilations due to missing components in Qiskit have been approved. The root cause of the remaining 627 miscompilations is under further investigation. 
}
\end{tcolorbox}

\subsection{Threats of Validity}

\textbf{Statevector Simulation Limitations.} Our analysis relies heavily on statevector simulations, which assume an idealized, noise-free quantum environment. While statevectors provide precise and complete representations of quantum program behavior, they do not account for decoherence, gate noise, or measurement errors present in real quantum hardware. Consequently, bugs or miscompilations that only manifest under noisy conditions may be overlooked. Future work should incorporate density matrix simulation or noisy backends to assess whether miscompilations persist in realistic settings.

\textbf{Randomness in Program Generation.} QSPE generates candidate programs by randomly instantiating holes, and the resulting program characteristics, such as control-flow shape or entanglement pattern, can vary widely across runs. Although we used a large sample set to mitigate randomness, the results may still be biased toward certain structural patterns that are more likely to occur due to the limited size of the set. This introduces a potential threat to the generalizability of our findings. More structured or guided generation strategies may reduce such bias.

\textbf{Test Oracle Granularity.} Our differential testing relies on the statevector as the oracle. However, some semantic divergences may remain undetected if the differences fall below the divergence threshold or if they do not manifest on the selected measurement basis. This means that some subtle miscompilations might be misclassified as equivalent programs. Incorporating multiple measurement bases or a formal equivalence check method could help strengthen the validity.

\textbf{Scalability Constraints.} Our experiments are conducted on 5-qubit programs for two main reasons. (1) Most existing quantum algorithms can be implemented with up to five qubits, so we consider scalability to be a less pressing issue at this stage. (2) Quantum circuit simulators require exponentially increasing computational time as the number of qubits grows beyond five, making larger-scale experiments computationally expensive. Larger circuits with more qubits and deeper gate counts may exhibit different miscompilation patterns or increase the likelihood of compiler-induced optimizations that lead to semantic shifts. Evaluating our method on scalable simulators or future fault-tolerant devices will be necessary to confirm these trends.

%% file: sec/6-ending.tex
\section{Related Work}
\label{sec:work}

\textbf{Differential testing} ~\cite{mckeeman1998differential,tang2020compiler} is a widely used strategy in traditional compiler validation, where the same input is compiled using multiple compilers to check for behavioral discrepancies. 
Classic tools like Csmith~\cite{yang2011finding} generate random C programs to uncover compiler bugs, while Quest~\cite{lindig2005random} targets argument passing and return values. Unlike random approaches, refactoring-based methods systematically transform programs to preserve semantics. Tools such as Orion~\cite{le2014compiler} and Orison~\cite{le2015finding} adapt EMI and mutation strategies to validate compilers, including GCC, LLVM, and OpenCL~\cite{lidbury2015many}. Zhang et al.~\cite{zhang2017skeletal} first propose extracting the skeleton (structure) of the program to generate a large set of program variants. However, due to the differences in platform features and program structures, the above classical approaches can hardly be applied to meet the new testing requirements of quantum programs.

\textbf{Quantum Testing and Verification.} Several efforts address correctness in quantum software. Zhao et al.~\cite{zhao2020quantum, zhao2025quantum} outline a lifecycle model for quantum software, while Ying et al.~\cite{ying2020symbolic} introduce formal reasoning via matrix-valued Boolean logic. Huang et al.~\cite{huang2019statistical} propose a logic-based framework for $\epsilon$-robust assertion checking in quantum programs. Tools like Proq~\cite{li2020projection} and QPMC~\cite{feng2015qpmc} support runtime assertion checking and model checking via quantum Markov chains. Ali et al.~\cite{ali2021assessing} introduce quantum input-output coverage and multiple test oracles, though their approach focuses on quantum programs rather than the full stack. Ye et al.~\cite{ye2023quratest} have revealed the impact of quantum states in the testing procedure of quantum programs. Regarding differential testing for quantum libraries, QDiff~\cite{wang2021qdiff}, Upbeat~\cite{hu2024upbeat}, and QuteFuzz~\cite{iwumbwe2025qutefuzz} have proposed various approaches to generate effective variants for detecting bugs.

\textbf{Verified Quantum Compilers.} Formal verification methods ensure compiler correctness by construction. For example, CertiQ~\cite{shi2019contract} verifies Qiskit transformations using a circuit equivalence calculus, while VOQC~\cite{hietala2021verified} builds on the CompCert infrastructure to formally verify quantum circuit optimizations. Other efforts, such as~\cite{smith2019quantum}, use quantum-equivalence structures, including QMDDs, to validate translations. Although these approaches emphasize provable correctness for compilers, quantum libraries face challenges from optimization and cross-platform transformations. To serve as a complementary technique that empirically identifies bugs across different quantum libraries with optimizations, QSPE makes enhancements and improvements to address the above challenges.

\section{Conclusion}
\label{sec:conclusion}

We propose QSPE, a practical approach for testing quantum libraries. QSPE extends the SPE approach to quantum platforms. In experiments, QSPE can quickly generate a large meanwhile valid set of program variants for libraries. Furthermore, by adopting the statevector-based validation method, false-positive reports can be reduced. Based on this, QSPE found 708 miscompilations, 81 of which have been approved by the developers of the Qiskit library.